\documentclass{llncs}

\usepackage{textcomp}
\usepackage{amsmath}
\usepackage{amsfonts}
\usepackage{amssymb}
\usepackage{graphicx,color}	
\usepackage{listings}

\lstnewenvironment{code}
    {\lstset{}%
      \csname lst@SetFirstLabel\endcsname}
    {\csname lst@SaveFirstLabel\endcsname}
\newcommand{\cpi}{c$\pi$}
\newcommand{\lbc}{$\mathcal{LBC}$}

\newcommand{\tl}[1]{\mathbf{#1}} 
\newcommand{\tlb}[2]{\mathbf{#1}_{#2}} 
\newcommand{\tli}[3]{\mathbf{#1}_{[#2,#3]}} 
\newcommand{\gtee}{\mathrel{\triangleright}} 
\newcommand{\pc}{\mathbin{\parallel}} 

\newcommand{\R}{\mathbb{R}} 
\newcommand{\Rpos}{\mathbb{R}_{\geq 0}}
\newcommand{\B}{\mathbb{B}}

\newcommand{\Pspace}{\mathbb{P}} 

\newcommand{\I}{\mathcal{I}} 

\newcommand{\Spec}{\mathcal{S}} 
\newcommand{\Proc}{\mathcal{P}} 

\newcommand{\gsep}{\;\; | \;\;} 
\newcommand{\bnf}{\;\; ::= \;\;} 

\newcommand{\sat}{\vDash} 
\newcommand{\nsat}{\nvDash}

\renewcommand{\leq}{\leqslant} 
\renewcommand{\geq}{\geqslant}

\usepackage{float}
\floatstyle{ruled}
\restylefloat{figure}

\begin{document}

\title{A More Sensitive Context}
\author{Christopher J.\ Banks \and Ian Stark}
\institute{Laboratory for Foundations of Computer Science\\ School of Informatics, University of Edinburgh, UK}

\maketitle

\begin{abstract}
  \emph{Logic of Behaviour in Context} (\lbc) is a spatio-temporal logic for
  expressing properties of continuous-state processes, such as biochemical
  reaction networks. {\lbc} builds on the existing \emph{Metric Interval
    Temporal Logic} (MITL) and adds a ``context modality'' that explores the
  behaviour of a system when composed with an external process.  {\lbc} models
  are terms of the \emph{Continuous $\pi$-Calculus}~(\cpi), a process algebra
  with continuous state space.
  
  Our previously published {\lbc} model-checking technique required examining
  many points along the behavioural trajectory of a process; and potentially
  computing further trajectories branching off at every such point.  This
  raised two difficulties: mixing temporal and spatial modalities could
  require computing a large number of trajectories, with costly numerical
  solution of differential equations; and might still fail to check
  intermediate values between discrete points on those trajectories.

  In this paper we make progress against both of these problems using
  techniques from signal temporal logic and from sensitivity analysis.
  Boolean signals aggressively compress trace information, allowing more
  efficient computation; and sensitivity analysis lets us reliably check
  formulae over a region by calculating a smaller number of sample
  trajectories.
\end{abstract}

\section{Introduction}

The Logic of Behaviour in Context (\lbc)~\cite{Banks2013} is a spatio-temporal
logic for expressing temporal and contextual properties of continuous state
processes (dynamical systems), such as biochemical reaction
processes. Temporal properties express behaviour over time and contextual
properties express behaviour in the presence of another process.  {\lbc}
equips a metric interval temporal logic with a \emph{context modality} which
asserts the behaviour of a process when composed with another process.  As
such, the logic requires a model with a suitable notion of process
composition; in this case we use the Continuous $\pi$-Calculus
(\cpi)~\cite{Kwiatkowski2008}.  We give a brief account of {\cpi} in
Section~\ref{cpi} which is sufficient for the reader to understand its use as
a model for {\lbc} and in Section~\ref{lbc} we give the syntax and semantics
of the logic.

In previous work~\cite{Banks2013} we gave various model checking algorithms
that can be used to verify the satisfaction of an {\lbc} formula over a {\cpi}
model.  Each of these algorithms takes an approximate approach to
model-checking continuous dynamical systems, using discrete simulation traces
computed by numerical solvers.  This a non-exact approach, but well founded in
practice for the efficient approximate model checking of continuous processes
(e.g.\ Antoniotti et al.~\cite{Antoniotti2003}, Fages and
Rizk~\cite{Fages2008}, or Nickovic and Maler~\cite{Nickovic2007}).  The
approach relies on the assumption that the numerical solver is suitably
precise, and that the discrete sample points of the simulation trace are
sufficiently dense to give a good approximation of the dynamics of the system.

Our method for model checking of the context modality relied on the same
assumption. To compute the satisfaction of a context modality we take sample
points along the trajectory of the model, compose with the new context
(process) at each point, and verify that the desired behaviour is observed on
trajectories computed from these.  A major problem with this approach is that
to be sure the sample points are dense enough for a good approximation one has
to numerically solve a large number of new trajectories, which is
computationally costly.  This approach also has the problem that there is no
way to ensure the property tested is not violated by trajectories starting
between the discrete sample points.

In this paper we address both of these problems: using Boolean signals to
compress traces, following Maler and Nickovic~\cite{Maler2004}; and
sensitivity analysis to reduce the number of sample trajectories, building on
Donz\'e and Maler~\cite{Donz2007}.

\section{Continuous \boldmath$\pi$-calculus}\label{cpi}

The continuous $\pi$-calculus (\cpi) was designed as a formal language for the
study of evolutionary variation in biochemical processes.  The canonical
reference for~{\cpi} is Kwiatkowski's thesis~\cite{Kwiatkowski2010}, but the
original language semantics were first published by Kwiatkowski and
Stark~\cite{Kwiatkowski2008}.

The language syntax is based on the~$\pi$-calculus of
Milner~\cite{Milner1999}, with some alterations and additions to better
support the description of biochemical models.  The usefulness
of~$\pi$-calculus style languages for modelling biochemical processes is well
established, having first been described by Regev et
al.~\cite{Regev2001,Regev2002}.

The description of a biochemical process in~{\cpi} is split into two levels:
\emph{species} and \emph{process}.  A \emph{species} in~{\cpi} is a
description of the behaviour and interaction capability of a biochemical
species.  This level is similar to a~$\pi$-calculus term.  A \emph{process}
in~{\cpi} is a real-indexed parallel composition of each of the species in the
biochemical process, representing a mixture with some initial concentration of
each species.

Kwiatkowski and Stark~\cite{Kwiatkowski2008} give full details, but for this
paper it is only necessary to understand that a {\cpi} process is a parallel
composition of species~$S$, each with an initial concentration~$c$.  The
species in a composition may, or may not, interact with each other.

\begin{definition}[Process]
  The set~$\Proc$ of \emph{{\cpi} processes} is defined by the following
  grammar:
  \begin{align*}
    \text{(Process) } P,Q \bnf& c \cdot S \gsep P \pc Q
  \end{align*}
  A \emph{process} may be a species~$S$, as defined in~\cite{Kwiatkowski2008},
  with initial concentration~$c\in\R$, or a composition (mixture) of species.
\end{definition}
A {\cpi} process has a compositional semantics in terms of real vector spaces,
from which an initial value problem, which gives the trajectory of a process,
can be extracted.  This is solved normally by a numerical simulator which
outputs a time series for each of the species in a process.  The space within
which we place these systems and the trajectories of their behaviour over time
is \emph{process space}.

\begin{definition}[Process space]
  The process space~$\Pspace$ is the vector space~$\R^{(\Spec^{\#})}$,
  where~$\Spec^{\#}$ is the set of prime species --- elementary species which
  cannot be broken down into a composition of two non-trivial species.
\end{definition}

\section{Logic of Behaviour in Context}\label{lbc}

The Logic of Behaviour in Context (\lbc)~\cite{Banks2013} arose from the
desire to define a logic for {\cpi} which would allow the classification of
the behaviour of a {\cpi} model.  It was clear that a temporal logic, and a
logic which allowed the expression of constraints on the real-valued
concentrations of {\cpi} species, was required.

However, particularly in biochemical systems, behaviour is often reasoned
about in terms of, not just the system itself, but the system's behaviour when
it is perturbed somehow.  We wish to be able to reason about the system's
behaviour in some external context.  Thus {\lbc} was conceived.

{\lbc} combines LTL($\R$) (see Calzone et al.~\cite{Calzone2006}) and MITL
(see Alur et al.~\cite{Alur1996}) with the addition of a \emph{context
  modality}.  Here LTL($\R$) is a temporal logic for properties of the
real-valued concentrations of biochemical species; and MITL adds concrete
times and time intervals to those temporal modalities.

The context modality~$(Q\gtee\psi)$ holds for a process~$P$ whenever~$\psi$
holds in the presence of a new process~$Q$.  This allows the expression of
behaviour in some given context.  For example, consider the assertion $(c\cdot
In)\gtee\tl{G}([Pr]<x)$.  In a biochemical context this could represent ``in
the presence of a concentration~$c$ of inhibitor~$In$ the concentration of
product~$Pr$ in the system always remains below~$x$''.

The context modality is based on the guarantee operator from Cardelli and
Gordon's spatial logic~\cite{Cardelli2000b}.  However, that guarantee takes an
arbitrary formula on the left hand side~$(\phi\gtee\psi)$ to give a formula
that holds for processes that when combined with any other process
satisfying~$\phi$ give a combination satisfying~$\psi$:
\[ 
P \models \phi\gtee\psi 
\quad\iff\quad \forall Q\;(Q\models\phi\implies P \pc Q\models\psi) \;.
\]
Model-checking a logic with this guarantee is hard because of the necessity to
quantify over all processes that satisfy an arbitrary formula.  Caires and
Lozes~\cite{Caires2006} give an account of the undecidability of spatial logic
with the guarantee.

The context modality, however, gives some of the power of guarantee in a more
computationally feasible form by specialising the left hand side to a specific
process:
\[ 
P \models Q\gtee\psi \quad\iff\quad Q \pc P\models\psi 
\] 
Nicola and Loreti~\cite{Nicola2005} take a similar approach with their logic
\texttt{MoMo} whose a production operator is based on guarantee.  However,
\texttt{MoMo} is defined for mobile processes with resources and locations,
modelled using a formalism based on shared tuple spaces, and the semantics of
the production operator is incompatible with the kind of continuous state
processes we address here.

\subsection{Syntax}

The syntax of {\lbc} follows MITL~\cite{Alur1996}, with propositional atoms
being inequalities between arithmetic combinations of real-valued
concentrations of species in the system and their time derivatives.  To this
we add the context modality~$Q\gtee\phi$.

\begin{definition}[{\lbc} formula] The set~$\Phi$ of~{\lbc} formulae~$\phi,\psi$ is defined by the following grammar:
  \begin{align*}
    \phi,\psi \bnf& \mathit{Atom} \gsep \phi\land\psi \gsep \phi\lor\psi \gsep \phi\Rightarrow\psi \gsep \lnot\phi \\
    \gsep & \phi\tlb{U}{I}\psi \gsep \tlb{F}{I}\phi \gsep \tlb{G}{I}\phi \gsep Q\gtee\phi \\
    \mathit{Atom} \bnf& \mathit{True} \gsep \mathit{False} \gsep \mathit{Val}\bowtie\mathit{Val} \\
    \mathit{Val} \bnf& v\in\R \gsep [S] \gsep [S]' \gsep \mathit{Val}\oplus\mathit{Val} \\
    \bowtie \bnf& > \gsep < \gsep \geq \gsep \leq \\
    \oplus \bnf& + \gsep - \gsep \times \gsep \div 
  \end{align*}\smallskip
  where relational operators~$\bowtie$ and arithmetic operators~$\oplus$ have
  the standard meaning, $[S]$ denotes the concentration of species~$S$, $[S]'$
  denotes the rate of change over time of the concentration of species~$S$,
  $Q$ is and {\cpi} process, and $I \subseteq \R^+$ is any non-negative time
  interval.
\end{definition}
We use the abbreviations~$\tl{U}$, $\tl{F}$, and~$\tl{G}$ to
denote~$\tli{U}{0}{\infty}$, $\tli{F}{0}{\infty}$, and~$\tli{G}{0}{\infty}$
respectively.  Likewise, for~$t \in \Rpos$, we write $\tlb{U}{t}$,
$\tlb{F}{t}$, and~$\tlb{G}{t}$ as abbreviations for $\tli{U}{0}{t}$,
$\tli{F}{0}{t}$, and~$\tli{G}{0}{t}$ respectively.

\subsection{Semantics}

We define the semantics of {\lbc} by its satisfaction relation~$\sat$, where
$P \sat \phi$ if and only if {\cpi} process $P$ satisfies formula~$\phi$.

\begin{definition}[Atomic propositions of a {\boldmath\cpi} process]
  \label{lbc-atom}
  The set $\mathit{Props}(P)$ of atomic propositions satisfied by a
  process~$P$ is defined by
  \begin{align*}
    \mathit{True}   &\in    \mathit{Props}(P) \\
    \mathit{False}  &\notin \mathit{Props}(P) \\
    v_1 \bowtie v_2 &\in \mathit{Props}(P) 
      \iff \mathit{value}(v_1,P)\bowtie\mathit{value}(v_2,P)
  \end{align*}
where the relational operators~$\bowtie$ are defined in the normal way and
\begin{align*}
  \mathit{value}(v,P) &= \text{$v$ for $v\in\R$}\\
  \mathit{value}([S_i],P) 
  &= \text{$c_i$ where $P = c_1\cdot S_1\pc\dots\pc c_n\cdot S_n$}\\
  \mathit{value}([S_i]',P) 
  &= \text{$c'_i$ where $dP/dt = c'_1\cdot S_1\pc\dots\pc c'_n\cdot S_n$}\\
  \mathit{value}(v_1\oplus v_2,P) 
  &= \mathit{value}(v_1,P)\oplus\mathit{value}(v_2,P)
\end{align*}
where arithmetic operations~$\oplus$ are similarly defined as normal.
\end{definition}

\begin{definition}[{\lbc} satisfaction relation]\label{lbc-sat}
  For~$P \in \Proc$, a {\cpi} process, and {\lbc} formulae~$\phi$ and~$\psi$ the satisfaction relation~$\sat$ is defined inductively as follows:
  \begin{center}
    \begin{tabular}{lll}
      $P \sat \mathit{Atom}$ & $\iff \mathit{Atom} \in \mathit{Props}(P)$\\
      $P \sat \phi \land \psi$ & $\iff P \sat \phi$ and $P \sat \psi$\\
      $P \sat \lnot \phi$ & $\iff P \nsat \phi$\\
      $P \sat \phi \tlb{U}{I} \psi$ & $\iff$ for some $t \in I, P^t \sat \psi$ and for all $t' \in [0,t], P^{t'} \sat \phi $\\
      $P \sat Q \gtee \phi$ & $\iff (Q \pc P) \sat \phi$
    \end{tabular}
  \end{center}
  where $Q \in \Proc$ is any {\cpi} process and process~$P^t$ is the state
  reached by process~$P$ after time~$t$; that is, the concentration of each
  species in~$P^t$ will be those present after~$P$ has run for time~$t$.  The
  notation~$P^t$ is shorthand for a function mapping
  ${\Proc\times\R^+\rightarrow\Proc}$, with $P^0 = P$.
\end{definition}
The remaining propositional connectives are derived as usual, together with
temporal modalities ${\tlb{F}{I}\phi \equiv \mathit{True}\tlb{U}{I}\phi}$ and
${\tlb{G}{I}\phi \equiv \lnot(\tlb{F}{I}(\lnot\phi))}$.

\begin{definition}[Duration of formula] 
  The \emph{duration} of a formula --- the length of time to which it refers
  --- is defined inductively:
  \begin{align*}
    |\mathit{Atom}| &= 0 &
    |\phi\land\psi| &= \max(|\phi|,|\psi|) &
    |\lnot\phi| &= |\phi| \\
    |Q\gtee\phi| &= 0 &
    |\phi\tli{U}{a}{b}\psi| &= \max(|\phi|,|\psi|) + b\;.
  \end{align*}
\end{definition}

\section{Signal checking}\label{oldsig}

In our earlier paper~\cite{Banks2013} we used a model-checking method based on
direct computation over time series from numerical simulation.  Here we refine
this technique using Boolean signals, inspired by Maler and
Nickovic~\cite{Maler2004}.

The key idea for signal-based temporal model checking is that the dynamics of
the model is represented as a set of Boolean signals.   Basic signals represent
whether an atomic proposition is satisfied at a given time; for each logical
operator we have a matching combinator for signals; and by applying these
recursively over the structure of a formula we compute its satisfaction
signal.

Simple formula satisfaction for a process is computed by taking the initial
value of the corresponding signal: $P\models\phi \iff
(s_\phi(0)=\mathit{True})$

Signals have a compact representation as sets of time intervals, in general
giving significant compression over time-series of data values.   The signal
combinators can be efficiently implemented for this representation, which
leads to substantial performance improvements over our earlier {\lbc}
implementation~\cite{Banks2013}.

\subsection{Signals}\label{signals}
Signals are constructed from the dynamics of a model.  A signal represents the satisfaction of a formula at any given time.  The set~$\mathit{Signal}$ is the set of finite length Boolean signals.
\begin{definition}[Finite length Boolean signal]\label{sigdef}
  A finite length Boolean signal~$s$ of length~$r$ is a function $s : [0,r)
  \rightarrow \B$.  A finite length Boolean signal has finite variability and,
  therefore, may be represented by a finite interval covering.  For a
  signal~$s$ with length~$r$ an interval covering is a
  sequence~$\I=I_1,I_2,\ldots$ of left-closed right-open intervals such that
  $\bigcup I_i = \I$ and $I_i \cap I_j = \emptyset$ for all $i \neq j$.  The
  minimal covering~$\I_s$ of the signal~$s$ is \emph{consistent} with the
  signal if $s(t)=s(t')$ for all~$t,t'$ in the same interval $I_i \in \I_s$.
  A covering~$\I'$ is a \emph{refinement} of~$\I$, denoted~$\I'\prec\I$, if
  for all~$I'\in\I'$ there exists~$I\in\I$ such that~$I'\subseteq I$.  The set
  of \emph{positive} intervals of~$s$ is $\I^+_s = \{I\in\I_s : s(I) =
  \mathit{True}\}$ and the set of \emph{negative} intervals is $\I^-_s = \I_s
  \setminus \I^+_s$.
\end{definition}
Signal checking relies on the conversion from the dynamics of the model to a
set of \emph{basic signals}, each of which represent the satisfaction of an
atomic proposition.

\subsubsection{Basic signals}

Basic signals are constructed from simulation traces of the form
$[(t_0,\vec{c_0}),\dots,(t_n,\vec{c_n})] \in \mathit{Trace}$, where each $t_i$
is a time point and~$\vec{c_i}$ is a vector of the species concentrations at
that time.  To construct the basic signals we use the following procedure:

\begin{enumerate}
\item Take each leaf~$\phi$ in the syntax tree of the formula; these are the
  atomic propositions of the form~$[A] \bowtie c$, $\mathit{True}$, or
  $\mathit{False}$.
\item For each~$\phi$ we construct a signal~$s_\phi$ as an interval
  covering~$\I$ of intervals $[t_0,t_1),[t_1,t_2),\ldots,[t_{n-1},t_n)$.
\item Each interval~$[t_i,t_{i+1})$ is in~$\I^+_{s_\phi}$ if the constraint
  in~$\phi$ is satisfied by the values in~$\vec{c_i}$, otherwise it is
  in~$\I^-_{s_\phi}$.
\end{enumerate}
This set of signals captures the value over time of all the atomic
propositions from a formula.

\subsubsection{Signal combinators}

For non-atomic formulae we have a set of signal combinators that take the
basic signals, apply a logical operation, and give the signal for the
satisfaction of a formula over time.  A signal~$s$ is constructed by computing
its covering intervals~$\I_s$; it is sufficient to compute the positive
intervals~$\I^+_s$ as the negative intervals~$\I^-_s$ are, by definition,
complimentary.

\begin{definition}[Boolean signal combinators~\cite{Maler2004}]\label{combinators}
  The signal combinators apply the logical connectives ($\lnot,\land$) and
  temporal modalities ($\tl{F},\tl{U}$) to signals ($s_\phi,s_\psi$) and are
  defined as follows:
  \begin{description}
  \item[$\lnot s_\phi$] \hfil\\
    Negation is a simple negation of the signal such that $\I^+_{s_{\lnot\phi}} = \I^-_{s_{\phi}}$.
  \item[$s_\phi \land s_\psi$] \hfill\\
    For conjunction we first compute a refinement of the coverings $\I^R_\phi
    \prec \I_\phi$ and $\I^R_\psi \prec \I_\psi$ such that $\I^R_\phi =
    \I^R_\psi$ and is the sequence of intervals $I^R_1,\ldots,I^R_n$.  The
    conjunction is then computed interval-wise such that $s_{\phi\land\psi} =
    s_\phi \land s_\psi$.  The minimal covering~$\I_{s_{\phi\land\psi}}$ is
    then computed by merging any adjacent intervals of the same Boolean value.
  \item[$\tli{F}{a}{b}s_\phi$] \hfill\\
    The temporal~$\tli{F}{a}{b}$ modality is computed by back-shifting the
    positive intervals.  $\I^+_{\tli{F}{a}{b}\phi}$ is constructed by taking
    each interval $I \in \I^+_{\phi}$ and computing its back-shifting $I
    \ominus [a,b] \cap \R^+$ where $[m,n) \ominus [a,b] = [m-b,n-a)$ and the
    intersection with~$\R^+$ eliminates any negative times.  The minimal
    covering~$\I_{\tli{F}{a}{b}\phi}$ is then computed by merging any adjacent
    intervals of the same Boolean value.
  \item[$s_\phi \tli{U}{a}{b} s_\psi$] \hfill\\
    The fundamental temporal~$\tli{U}{a}{b}$ modality can be computed on the
    basis that $\phi\tli{U}{a}{b}\psi \iff \phi \land \tli{F}{a}{b}(\phi \land
    \psi)$ when~$s_\phi$ is a unitary signal.  A signal~$s$ is unitary
    if~$\I^+_s$ is a singleton.  So if~$s_\phi$ is unitary and it holds
    at~$t_1$ and~$t_2$ then it must hold for the whole interval~$[t_1,t_2]$.
    For the case where~$s_\phi$ is not unitary we can decompose it into a set
    of unitary signals $\{s_\phi^1,\ldots,s_\phi^n\}$ and compute, for each~$i
    \in [1,n]$:
    \[ 
    s_{\phi\tli{U}{a}{b}\psi}^i = s_\phi^i \land \tli{F}{a}{b}(s_\phi^i
    \land s_\psi) 
    \] 
    The signal is then recomposed to give:
    \[ 
    s_{\phi\tli{U}{a}{b}\psi} = \bigvee_{i=1}^n s_{\phi\tli{U}{a}{b}\psi}^i 
    \]
  \end{description}
\end{definition}

\subsubsection{Context modality signal}
Our problem is how to compute the signal for the context
modality~$Q\gtee\phi$.  The positive intervals of the
signal~$\I^+_{s_{Q\gtee\phi}}$ must represent the times at which if~$Q$ is
composed with the model then~$\phi$ is satisfied.  To compute this we must
choose a finite number of arbitrary time points at which to introduce~$Q$ to
the model and compute the satisfaction of~$\phi$.  The problem lies in how to
choose these time points.

An initial solution is to choose the same time points as in the original
trace; that is if we are checking $P \models q\gtee\phi$ then we use the same
time points as in the trace for~$P$.  The assumption here is that if the
chosen time points for~$P$ were sufficiently dense then they will be
sufficiently dense for~$Q\gtee\phi$.

\begin{definition}[Context modality signal]\label{oldcontext}
  The context modality signal is constructed as follows.  To compute a signal
  for $P\models Q\gtee\phi$, for each time point~$t$ in the originally
  computed simulation trace we compute a new process $P^t \pc Q$.  Each of
  these new processes is solved numerically to get a trace and we recursively
  apply the signal checking procedure to find whether or not~$\phi$ holds for
  each of these processes.  The Boolean result from each process at time~$t$
  is the value of the signal for the interval~$[t,t')$ where~$t'$ is the time
  of the next point.
\end{definition}

The use of signal checking has, in a preliminary implementation, shown good
improvements in performance for checking the temporal fragment of the logic.
However, the focus of this paper is to show that improvements can also be made
in the following technical issues.

\subsection{Outstanding issues}\label{problems}

There are two main issues with this approach to checking $P \sat Q \gtee
\phi$.  If we wish to check a temporal nested context modality, say $P \sat
\tli{G}{0}{t} (Q \gtee \phi)$, then taking sample points on the original
trajectory of~$P$ and checking $P^{t'} \pc Q \sat \phi$ for each sample
point~$t'\in[0,t]$ does not necessarily ensure that $P^{t''} \pc Q \sat \phi$
holds for~$t''$ between sample points.  We have simply worked on the
assumption that with sufficiently dense sample points we will have great
enough coverage to have a reasonable degree of confidence in the result.  This
seems a reasonable assumption to make, given that this applies to all
applications of numerical simulation.

The second issue is that having a very large set of sample points means that
we have to make a very large number of calls to the solver in order to
solve~$P^t \pc Q$ for each sample point~$t$.   Experience with real examples
has shown this is easily the most significant factor in the computational cost
of model-checking {\lbc}.

\section{Context checking with sensitivity}\label{sens}
Our new method addresses both of the problems outlined in
Section~\ref{problems}.  The method potentially reduces calls to the ODE
solver whilst also improving the coverage of context introduction between the
original sample points.  The key to the method comes from a study of
sensitivity analysis for safety properties by Donz\'e and
Maler~\cite{Donz2007}.  Sensitivity analysis is used to systematically check a
system with uncertain initial conditions; the use of sensitivity analysis
ensures that a few discrete simulations cover a continuous space of initial
conditions.  Here we can apply the same principle allowing $P^t \pc Q \sat
\phi$ to be checked for all~$t$, because the difference between $P \pc Q$ and
$P^t \pc Q$ is just a change in initial conditions.  This initial condition
space is potentially large, but we can find a sufficient and finite set of
samples using an adaptation of Donz\'e and Maler's technique; thus potentially
reducing the number of calls made to the ODE solver.

\subsection{Expansion function}
If we take the set~$X_0$ to be the set of initial conditions for the dynamical
system corresponding to the processes $P^{t_0} \pc Q,\ldots,P^{t_n} \pc Q$,
where~$Q$ is introduced to~$P$ at times in~$[t_0,t_n]$, then we
take~$\xi_{x_0}(t)$ to be the trajectory which is the unique solution for some
initial condition~$x_0 \in X_0$.  The set of states reachable within time~$t$
from any initial condition is denoted $\mathtt{reach}_{\leq t}(X_0)$ and the
set of states reachable at exactly time~$t$ is denoted
$\mathtt{reach}_{=t}(X_0)$.  Let~$d$ be a distance metric between points and
let it extend to points and sets thus $d(x,X) = \inf_{y \in X} (d(x,y))$.

The set~$\mathcal{B}_\delta(x)$ is the $\delta$-ball around point~$x$ and the
set $\mathcal{B}_\delta(X)$ is the $\delta$-ball around set~$X$ thus
$\mathcal{B}_\delta(X) = \bigcup_{x \in X}\mathcal{B}_\delta(x)$.  Donz\'e and
Maler show that it is possible to find the ball which tightly
over-approximates $\mathtt{reach}_{=t}(\mathcal{B}_\delta(x_0))$ by means of
the \emph{expansion function}.  Therefore it is possible to construct a ``flow
tube'' around a trajectory which tightly over-approximates the reachable set.

\begin{definition}[Expansion function~\cite{Donz2007}]
  Given~$x_0 \in X_0$ and some~$\epsilon>0$ the \emph{expansion function} of
  trajectory~$\xi_{x_0}$, denoted by $\Xi_{x_0,\epsilon}:\R^+\rightarrow\R^+$
  maps~$t$ to the smallest non-negative number~$\delta$ such that all
  trajectories with initial state in~$\mathcal{B}_\epsilon(x_0)$ reach a point
  in~$\mathcal{B}_\delta(\xi_{x_0}(t))$ at time~$t$:
  \[
  \Xi_{x_0,\epsilon}(t) = \sup_{d(x_0,x)\leq\epsilon} d(\xi_{x_0}(t), \xi_x(t))
  \]
\end{definition}
The value of the expansion function is the radius of the tightest ball around
the reachable set from the $\epsilon$-ball around the initial condition.  The
key here is that if we take the initial set to be the ball which tightly
bounds our possible initial conditions then we can compute an
over-approximation of the reachable set and therefore prove that we do not
reach a state where some~$\phi$ holds.  Donz\'e and Maler~\cite{Donz2007} show
that $\Xi_{x_0,\epsilon}(t)$ can be computed via \emph{sensitivity to initial
  conditions} $\frac{\partial \xi_{x_0}}{\partial x_0}(t)$, which is commonly
implemented by numerical solvers.  The error in the numerical approximation is
quadratic in~$\epsilon$~\cite{Donz2007}.  Therefore, for our algorithm we have
a parameter~$\theta$ which is the maximum initial ball radius; if the ball
around our initial conditions is greater than~$\theta$ then we refine until
the initial balls are all smaller than~$\theta$.

Donz\'e and Maler then have a scheme for refining the over-approximation for
an arbitrary set of initial conditions in~$\R^n$.  However here our initial
conditions are less general and so, in Section~\ref{sig}, we give a more
specific method used for computing the signal of a context modality.

\subsection{Application to checking \boldmath\lbc}\label{sig}
We can apply the principle of computing traces of flow tubes, where necessary,
instead of traces of trajectories.  Some preliminary definitions follow.

The set~$B$ is the set of balls in~$\Pspace$ and for~$\beta\in B$: $c(\beta)$
is the centre point of the ball and~$r(\beta)$ is the radius.  The
set~$\mathit{Tube}$ is the set of flow tube traces in~$\Pspace$ where a flow
tube trace is of the form $(t_0,\beta_0),\ldots,(t_n,\beta_n)$,
where~$t_i\in\R^+$ is a time point and~$\beta_i\in B$.  The ball~$\beta\pc P$
is the ball~$\beta$ translated by the process vector~$P$; that is, $\beta$
translated by the concentration value for each dimension (species) in $P$.

\begin{definition}[Trace and flow tube trace]\\
  We have a function
  ${\mathtt{trace}:\Pspace\times\R^+\times\R^+\rightarrow\mathit{Trace}}$,
  where $\mathtt{trace}(P,t,\rho)$ gives the trace of process~$P$ up to
  time~$t$ with resolution~$\rho$ using numerical simulation.  We also have a
  function ${\mathtt{tube}:B\times\R^+\times\R^+\rightarrow\mathit{Tube}}$,
  where $\mathtt{tube}(\beta,t,\rho)$ gives the flow tube trace from the
  ball~$\beta$ for time~$t$ with resolution~$\rho$; each ball in the flow tube
  trace is given by $\Xi_{c(\beta),r(\beta)}(t_i)$ for each~$t_i$ up to~$t$
  such that~$t_{i+1}-t_i=\rho$.
\end{definition}
The set~$\B_\bot$ is $\{\mathit{True},\mathit{False},\bot\}$ where~$\bot$ is
of ``uncertain'' Boolean value.  The set~$\mathit{Signal}_\bot$ is the set of
finite length~$\B_\bot$ signals, signals with uncertainty, defined as follows:
\begin{definition}[Finite length~\boldmath$\B_\bot$ signal]
  A finite length~$\B_\bot$ signal~$s$ of length~$r$ is a function $s : [0,r)
  \rightarrow \B_\bot$.  A finite length~$\B_\bot$ signal has finite
  variability and, therefore, may be represented by a finite interval
  covering.  The interval covering, minimal covering, and definitions of
  consistency and refinement are the same as in Definition~\ref{sigdef}.

  The set of \emph{positive} intervals of~$s$ is ${\I^+_s = \{I\in\I_s : s(I)
    = \mathit{True}\}}$, the set of \emph{uncertain} intervals of~$s$ is
  ${\I^\bot_s = \{I\in\I_s : s(I) = \mathit{\bot}\}}$, and the set of
  \emph{negative} intervals is ${\I^-_s = \I_s \setminus (\I^+_s \cup
    \I^\bot_s)}$.
\end{definition}

\begin{definition}[\boldmath$\B_\bot$ signal combinators]
  \label{def:signal3combinators}
  The signal combinators apply the logical connectives ($\lnot,\land$) and
  temporal modalities ($\tl{F},\tl{U}$) to signals ($s_\phi,s_\psi$) and are
  defined as follows:
  \begin{description}
  \item[$\lnot s_\phi$] \hfil\\
    Negation is a simple negation of the signal such that
    $\I^+_{s_{\lnot\phi}} = \I^-_{s_{\phi}}$ and $\I^\bot_{s_{\lnot\phi}} =
    \I^\bot_{s_{\phi}}$.
  \item[$s_\phi \land s_\psi$] \hfill\\
    For conjunction we first compute a refinement of the coverings $\I^R_\phi
    \prec \I_\phi$ and $\I^R_\psi \prec \I_\psi$ such that $\I^R_\phi =
    \I^R_\psi$ and is the sequence of intervals $I^R_1,\ldots,I^R_n$.  The
    conjunction is then computed interval-wise such that $s_{\phi\land\psi} =
    s_\phi \land s_\psi$ where $s_{\phi\land\psi} = \bot$ when either
    $s_{\phi} = \bot$ or $s_{\psi} = \bot$.  The minimal
    covering~$\I_{s_{\phi\land\psi}}$ is then computed by merging any adjacent
    intervals of the same value.
  \item[$\tli{F}{a}{b}s_\phi$] \hfill\\
    The temporal~$\tli{F}{a}{b}$ modality is computed by back-shifting the
    positive and uncertain intervals.  $\I^+_{\tli{F}{a}{b}\phi}$ is computed
    by taking each interval $I \in \I^+_{\phi}$ and computing its
    back-shifting $I \ominus [a,b] \cap \R^+$.  In the same way
    $\I^\bot_{\tli{F}{a}{b}\phi}$ is computed and where any interval overlaps
    with an interval in~$\I^+_{\tli{F}{a}{b}\phi}$ the overlapping portion of
    the interval is removed from $\I^\bot_{\tli{F}{a}{b}\phi}$.  The minimal
    covering~$\I_{\tli{F}{a}{b}\phi}$ is then computed by merging any adjacent
    intervals of the same Boolean value.
  \item[$s_\phi \tli{U}{a}{b} s_\psi$] \hfill\\
    The fundamental temporal~$\tli{U}{a}{b}$ modality can be computed on the
    basis that $\phi\tli{U}{a}{b}\psi \iff \phi \land \tli{F}{a}{b}(\phi \land
    \psi)$ when~$s_\phi$ is a unitary signal.  A signal~$s$ is unitary
    if~$\I^+_s\cup\I^\bot_s$ is a singleton.  So if~$s_\phi$ is unitary and it
    has a value at~$t_1$ and~$t_2$ then it must hold that value for the whole
    interval~$[t_1,t_2]$.  For the case where~$s_\phi$ is not unitary we can
    decompose it into a set of unitary signals $\{s_\phi^1,\ldots,s_\phi^n\}$
    and compute, for each~$i \in [1,n]$:
    \[
    s_{\phi\tli{U}{a}{b}\psi}^i 
    = s_\phi^i \land \tli{F}{a}{b}(s_\phi^i \land s_\psi) 
    \] 
    The signal is then recomposed to give: 
    \[ 
    s_{\phi\tli{U}{a}{b}\psi} 
    = \bigvee_{i=1}^n s_{\phi\tli{U}{a}{b}\psi}^i 
    \]
  \end{description}
\end{definition}
Our sensitive model checker for {\lbc} over {\cpi} processes is defined by
four mutually recursive functions.  Function $\mathtt{sat}$ is the top-level
function which computes whether or not a process satisfies an {\lbc} formula.
Function $\mathtt{satB}$ computes whether a ball satisfies a formula: giving
$\mathit{True}$, $\mathit{False}$ or~$\bot$ according to whether every point
of the ball satisfies the formula, none do, or only some.  The function
$\mathtt{signal}$ computes the formula satisfaction signal along a process
trajectory.  Finally, $\mathtt{signalT}$ computes a satisfaction signal along
a flow tube trace.  As with $\mathtt{satB}$, this signal is three-valued
according to whether the cross-section of the flow tube at each trace instant
lies within, outside, or partially within the region satisfying the formula.

The algorithms for our sensitive model checker have two control parameters: a
time resolution~$\rho$ as the step size for traces and flow tube traces; and
$\theta$, the ball radius within which we use the expansion function to
extrapolate a flow tube.

\begin{definition}[$\mathtt{sat}$]
  Function~$\mathtt{sat}:\Pspace\times\Phi\rightarrow\B$ is computed
  recursively as follows:
  \begin{align*}
    \mathtt{sat}(P,\mathit{Atom})     &= \mathit{Atom}\in\mathit{Props}(P)\\
    \mathtt{sat}(P,\phi\land\psi)    
      &= \mathtt{sat}(P,\phi)\land\mathtt{sat}(P,\psi)\\
    \mathtt{sat}(P,\lnot\phi)        &= \lnot(\mathtt{sat}(P,\phi))\\
    \mathtt{sat}(P,(Q\gtee\phi)) &= \mathtt{sat}(P \pc Q,\phi)\\
    \mathtt{sat}(P,\phi\tli{U}{a}{b}\psi) 
    &= (\mathtt{signal}(P,|\phi\tli{U}{a}{b}\psi|,\phi\tli{U}{a}{b}\psi))(0)\;.
  \end{align*}
\end{definition}
Satisfaction of a non-temporal formula is straightforward and can be
determined directly from the initial conditions.  Even a context formula, if
its subformula is non-temporal, requires only a process composition and an
inspection of initial conditions.  However, for a temporal formula we compute
the signal of its satisfaction over time; and then take the initial value of
that signal.  This leads us to the next function required.

\begin{definition}[$\mathtt{signal}$]
  Function~$\mathtt{signal}:\Pspace\times\R^+\times\Phi%
  \rightarrow\mathit{Signal}$ is computed recursively as follows, using the
  Boolean signal combinators from Definition~\ref{combinators}:
  \begin{align*}
    \mathtt{signal}(P,t,\mathit{Atom}) 
    &= \mathtt{basicSignal}(P,t,\mathit{Atom})\\
    \mathtt{signal}(P,t,\phi\land\psi) 
    &= \mathtt{signal}(P,t,\phi) \land \mathtt{signal}(P,t,\psi)\\
    \mathtt{signal}(P,t,\lnot\phi) 
    &= \lnot(\mathtt{signal}(P,t,\phi))\\
    \mathtt{signal}(P,t,\phi\tli{U}{a}{b}\psi) 
    &= \mathtt{signal}(P,t,\phi) \tli{U}{a}{b} \mathtt{signal}(P,t,\psi)\\
    \mathtt{signal}(P,t,Q\gtee\phi) 
    &= \mathtt{contextSignal}(P,t,Q,\phi)
  \end{align*}
  where $\mathtt{basicSignal}$ and $\mathtt{contextSignal}$ are defined below.
  \begin{description}
  \item[$\mathtt{basicSignal}:$]$\Pspace\times\R^+\times\mathit{Atomic}%
    \rightarrow\mathit{Signal}$\\
    Here $\mathtt{basicSignal}(P,t,\mathit{Atom})$ gives the finite signal $s$
    such that for each $(t_i,\vec{c_i})\in\mathtt{trace}(P,t,\rho)$:
    $s([t_i,t_i+\rho)) = (\mathit{Atom}\in\mathit{Props}(\vec{c_i}))$.
  \item[$\mathtt{contextSignal}:$]
    $\Pspace\times\R^+\times\Pspace\times\Phi\to\mathit{Signal}$\\
    Function $\mathtt{contextSignal}(P,t,Q,\phi)$ computes a signal~$s$ as
    follows.  Take set~$X$ of all process states in the trace $\tau =
    \mathtt{trace}(P,t,\rho)$, let~$\beta_X$ be the minimum bounding ball
    around~$X$, and let~$I$ be the interval~$[t_0,t_n+\rho)$ including time
    points~$t_0,\dots,t_n$ of~$\tau$.  If $\mathtt{satB}(\beta_X\pc Q,\phi)$
    is either $\mathit{True}$ or $\mathit{False}$ then
    ${s(I)=\mathtt{satB}(\beta_X\pc Q,\phi)}$ defines our signal~$s$.  If not,
    and $\tau$ contains only one time point $(t,\vec{c})$, then take ${s(I)=
      \mathtt{sat}(\vec{c}\pc Q,\phi)}$.  Finally, if $\mathtt{satB}$
    gives~$\bot$ and $\tau$ contains multiple points, then bisect~$\tau$ and
    repeat the procedure for each new $\tau$, $X$ and~$I$.
  \end{description}
\end{definition}
The signal for an atomic proposition is computed directly from the simulation
trace, interpolating between time points.  The signal for non-atomic formulae
other than context modalities is computed by using the appropriate signal
combinators.

The difficult case is for context modalities.  For this we compute a trace as
for atomic propositions, translate it by the context~$Q$, and then test a
bounding ball around all points in the trace.  If this is inconclusive we
repeatedly refine until we find a set of balls which give a conclusive result.
In the worst case this means checking individual points of the trace --- as we
did in our earlier methods.  However, in any other case, we may save
computation by checking a whole set of points together in a single ball.

That, however, requires a function to compute satisfaction across a ball.
\begin{definition}[$\mathtt{satB}$]
  The three-valued function~$\mathtt{satB}:B\times\Phi\rightarrow\B_\bot$ for
  satisfiability of a formula across a ball is computed recursively:
  \begin{align*}
    \mathtt{satB}(\beta,\mathit{Atom}) &= 
    \begin{cases}
      \mathit{True} & \beta \subseteq
      \{x\in\Pspace\mid\mathit{Atom}\in\mathit{Props}(x)\} \\
      \mathit{False} &
      \beta\cap\{x\in\Pspace\mid\mathit{Atom}\in\mathit{Props}(x)\} =
      \emptyset \\
      \bot & \text{otherwise}
    \end{cases}\\
    \mathtt{satB}(\beta,\phi\land\psi) &= 
    \begin{cases}
      \mathit{True} & (\mathtt{satB}(\beta,\phi)=\mathit{True}) 
      \land (\mathtt{satB}(\beta,\psi)=\mathit{True})\\
      \bot & (\mathtt{satB}(\beta,\phi)=\bot) 
      \lor (\mathtt{satB}(\beta,\psi)=\bot) \\
      \mathit{False} & \text{otherwise}
    \end{cases}\\
    \mathtt{satB}(\beta,\lnot\phi) &= 
    \begin{cases}
      \mathit{True}  & \mathtt{satB}(\beta,\phi) = \mathit{False}\\
      \bot           & \mathtt{satB}(\beta,\phi) = \bot\\
      \mathit{False} & \mathtt{satB}(\beta,\phi) = \mathit{True}
    \end{cases}\\
    \mathtt{satB}(\beta,Q\gtee\phi) &= \mathtt{satB}(\beta\pc Q,\phi) \\
    \mathtt{satB}(\beta,\phi\tli{U}{a}{b}\psi) &=
    \begin{cases}
      (\mathtt{signalT}(\mathtt{tube}(\beta,|\phi\tli{U}{a}{b}\psi|),\phi\tli{U}{a}{b}\psi))(0) & r(\beta) \leq \theta \\
      \bot & \text{otherwise}
    \end{cases}
  \end{align*}
\end{definition}
For atomic propositions, logical combinations, and the context modality, we
need only determine whether a ball lies entirely inside or outside the region
defined by the formula; with some refinement for combinations with the mixed
value~$\bot$.  For temporal formulae, if the ball has a radius too large for
reliable extrapolation by the expansion function, then we return~$\bot$.
Note, though, that where $\mathtt{satB}$ has been called from the within the
loop of $\mathtt{contextSignal}$ this will immediately trigger bisection and
further calls to $\mathtt{satB}$ over a smaller region.  Finally, to compute
validity of a temporal formula over a ball of radius $\theta$ or less, we
compute the signal over a flow tube trajectory and take its initial value.

That, of course, requires that we compute a signal over the tube trace.
\begin{definition}[$\mathtt{signalT}$]
  The function
  $\mathtt{signalT}:\mathit{Tube}\times\Phi\rightarrow\mathit{Signal}_\bot$ is
  defined recursively as follows:
  \begin{align*}
    \mathtt{signalT}(T,\mathit{Atom}) 
    &= \mathtt{basicSignalT}(T,\mathit{Atom}) \\
    \mathtt{signalT}(T,\phi\land\psi) 
    &= \mathtt{signalT}(T,\phi) \land \mathtt{signalT}(T,\psi)\\
    \mathtt{signalT}(T,\lnot\phi) &= \lnot (\mathtt{signalT}(T,\phi))\\
    \mathtt{signalT}(T,\phi\tli{U}{a}{b}\psi) 
    &= \mathtt{signalT}(T,\phi) \tli{U}{a}{b} \mathtt{signalT}(T,\psi)\\
    \mathtt{signalT}(T,Q\gtee\phi) &= \mathtt{contextSignalT}(T,Q,\phi)\;.
  \end{align*}
  This uses the operations on three-valued signals from
  Definition~\ref{def:signal3combinators} and auxiliary functions
  $\mathtt{basicSignalT}$ and $\mathtt{contextSignalT}$.
  \begin{description}
  \item[$\mathtt{basicSignalT}:$]
    $\mathit{Tube}\times\mathit{Atomic}\to\mathit{Signal}_\bot$ \\
    For this $\mathtt{basicSignalT}(T,\mathit{Atom})$ is the finite
    three-valued signal~$s$ where for each $(t_i,\beta_i)\in T$ we have
    ${s([t_i,t_i+\rho)) = \mathtt{satB}(\beta_i,\mathit{Atom})}$.
  \item[$\mathtt{contextSignalT}:$]
    $\mathit{Tube}\times\Pspace\times\Phi\to\mathit{Signal}_\bot$\\
    Function $\mathtt{contextSignalT}(T,Q,\phi)$ computes a three-valued
    signal as follows.  Let $W$ be the set of all balls in the tube trace~$T$,
    let $\beta_W$ be the minimum bounding ball around~$W$, and let $I$ be the
    interval~$[t_0,t_n+\rho)$ including time points~$t_0,\dots,t_n$ of~$T$.
    If $\mathtt{satB}(\beta_W\pc Q,\phi)$ is either $\mathit{True}$ or
    $\mathit{False}$ then we have our signal defined by
    ${s(I)=\mathtt{satB}(\beta_W\pc Q,\phi)}$.  Otherwise, if the tube trace
    $T$ contains only a single time point~$(t,\beta)$ then
    ${s(I)=\mathtt{satB}(\beta\pc Q,\phi)}$.  Finally, if $\mathtt{satB}$
    gives~$\bot$ and $T$ contains multiple time points, then bisect~$T$ and
    repeat the procedure for each new tube~$T$, set of balls~$W$ and
    interval~$I$.
  \end{description}
\end{definition}
Computation of a three-valued signal for a flow tube is very similar to that
for the boolean signal of a trajectory, except that we deal in balls rather
than points.  However, where in $\mathtt{contextSignal}$ we group points into
a ball to send to $\mathtt{satB}$, here we group balls into their bounding
ball, and still need only call $\mathtt{satB}$ --- closing off the mutually
recursive definition of our four functions.

The key advance here is that by using sensitivity calculations to drive the
expansion function, we may replace multiple calls to a numerical solver to
repeatedly compute traces with a single call to compute a flow tube.  The full
power of this is engaged when we have alternation of temporal and contextual
modalities, and may need to calculate trajectories starting at every point
along a spinal translated trajectory.  The sensitive model-checker groups
trajectories starting within $\theta$ of each other together into a single
flow tube and a single signal.  In the best case, we may tremendously reduce
the number of calls to the numerical solver and the number of signals to
compute; in the worst case, subdivision leads us to the previous algorithm of
a trace at every point.

This reduction in computation addresses the second problem
of~\S\ref{problems}.  Our algorithm also begins to address the first problem
there, of intermediate values between discrete points on a trace.  When
$\mathtt{satB}$ evaluates a formula of {\lbc} across a ball including several
points of a trajectory, it also does so for all values between them.  This
extends to flow tubes, too, surrounding all trajectories that start in their
initial ball.  This considerably extends the earlier, purely pointwise,
algorithm.  However, it is not necessarily complete as a very volatile
trajectory may conceivably range outside this enclosing ball in between trace
points.

\section{Conclusion}

In this paper we have shown ways to address two problems associated with using
numerical ODE solvers for model-checking contextual properties of continuous
dynamical systems.  We use methods from signal logics and from sensitivity
analysis.

The first problem was that of computational cost, arising from checking a very
large number of time-points along execution traces from the numerical solver.
The use of signal-logic methods reduces the overhead in working with large
traces, and for the context modality in particular we have identified ways to
use sensitivity analysis in numerical solvers to check a covering collection
of small regions rather than many individual points.  This aims to
substantially reduce the number of subsequent calls to the solver to check
nested spatial modalities.

The second problem was related to the correctness of the model checker when
working with discrete-time traces.  By computing with flow tubes around
trajectories that cover a computed set of initial conditions, we have a
stronger guarantee that we have not missed additional trajectories from
intermediate time points.  However, it is still possible to miss initial
conditions that lie outwith the bounding ball of the discrete time points on a
trace.  As would be expected, the correctness of this model-checking also
remains dependent on the accuracy of the numerical solver.

We have done some implementation, validating these approaches.  In particular,
we have a full implementation of the signal-logic techniques of~\S\ref{oldsig}
in our existing {\lbc}-checker, and can confirm a substantial performance
improvement from the highly-compressed signal representation of traces.  We
have not yet implemented the methods using expansion functions and sensitivity
analysis of~\S\ref{sens}.

\subsection*{Acknowledgements}

We are very grateful to Luca Bortolussi for suggesting the use of signals, and
introducing us to the work of Maler et al.; and to others in the Edinburgh
PEPA group.  This work was supported by the Laboratory for Foundations of
Computer Science, and by the Engineering and Physical Sciences Research
Council under grant number EP/P50550X/1.

\bibliographystyle{acm}
\bibliography{/home/chris/library} 

\end{document}